# Tailor-made Metasurfaces for Scattering Control

**M. Tsukerman[1], K. Grotov[1], A. Mikhailovskaya[1*], P. Bezrukov[1], S. Geyman[1], A. Kharchevskii[1], A. Maximenko[1], I. Yusupov[1], V. Bobrovs[1], and P. Ginzburg[1]**

[1] Institute of Telecommunications, Riga Technical University, Riga, 1048, Latvia
*Corresponding author. Email: mikhailovskaya1994@gmail.com

**Abstract:** Reducing electromagnetic scattering from an object has always been a task, inspiring efforts across disciplines such as materials science and electromagnetic theory. The pursuit of electromagnetic cloaking significantly advanced the field of metamaterials, yet achieving broadband, conformal cloaking for complex, non-trivial objects remains an unresolved challenge. Here, we introduce the concept of *tailored metasurfaces*: evolutionarily designed, aperiodic structures optimized to suppress scattering from arbitrarily shaped objects by accounting for their geometry, including resonant and large-scale features. Specifically, we demonstrate broadband scattering suppression of more than 20–30 dB over a ~20% fractional bandwidth for a variety of test objects, including randomly distributed wire meshes, spheres, and polyhedra. Beyond the conceptual advancement, the demonstrated performance positions this approach as a valuable technological solution for a range of wireless applications, including scenarios where interference from nearby objects degrades wireless channel performance. The evolutionary optimization framework enables efficient customization of electromagnetic devices, facilitating the development of high-performance structures that address complex technological demands in wireless systems, including interference mitigation, radar cross-section control, and signal integrity enhancement in next-generation networks.

## Introduction

The manipulation of electromagnetic wave propagation, facilitated by structured media, has long been a core focus of fundamental and applied research. To expand the available degrees of freedom within structured media, the concept of metamaterials was introduced[1–6], followed by their two-dimensional counterparts, metasurfaces, which address additional aspects and constraints such as reduced form factor, fabrication simplicity, and others[7–9]. One peculiarity introduced by the field of metamaterials is the negative refractive index[10–13] which was later proposed as a method for achieving electromagnetic cloaking[2,14,15]. This concept, aimed at rendering objects invisible to electromagnetic systems such as radars, has been proposed and demonstrated in various configurations and spectral ranges. The pioneering demonstration in the GHz frequency band was done with a cloaking device composed of an array of electric and magnetic resonances[14]. This arrangement creates an effective negative index metamaterial that guides electromagnetic waves around objects, preventing target-specific interactions and rendering them invisible. While this groundbreaking experiment marked a significant milestone in the field, it also revealed the inherent fundamental limitations of the approach. The primary limitations include the inherently narrow bandwidth due to the resonant nature of the effect and the bulky implementation required to precisely control the propagation phase for effective scattering cancellation. These challenges are critical considerations in advancing the practicality and efficiency of metamaterial-based cloaking technologies. However, in many applications, a weaker property of scattering suppression provides a reasonable compromise between reliable performance and practicality. For example, Mantle cloaking utilizes a specially designed thin layer to cover an object. This layer cancels out the waves scattered by the target by interfering destructively with them, thus reducing the object's visibility[16]. However, the primary efforts in this direction concentrate on relatively simple shapes (e.g., flat surfaces, cylinders, and spheres); thus, optimizing periodic arrays of resonators (metasurfaces) comes as a reasonable solution. However, in practical applications, objects targeted for scattering suppression often have complex shapes and are made from diverse materials, including metals and dielectrics. The varied shapes and material properties pose significant challenges compared to simpler geometries, inspiring the development of universal or adaptable scattering suppression solutions. An additional factor to consider is the bandwidth, the angle of incidence, and polarization. The bandwidth aspects become primarily challenging in the context of resonant structures. Considering all those aspects, the problem of scattering suppression from an arbitrary object becomes a subject of multi-objective optimization in a relatively large search space. It is worth briefly mentioning the field of absorbing materials, which are widely employed in radar and wireless applications, such as mitigating electromagnetic interference and multipath effects caused by nearby objects[17]. One example includes absorbers deployed on walls in an anechoic chamber. Despite their exceptional performance, this approach requires substantial wave propagation in the absorber, making it inherently bulky. Moreover, addressing resonant interaction regimes, especially when object features are wavelength-comparable, can be challenging with such methods, potentially limiting their universality.

Here, we present the concept of a tailor-made metasurface cover, engineered using an evolutionary algorithm to suppress electromagnetic scattering from arbitrarily shaped objects across a broad frequency range. This structure is to be attached in front of a target object to suppress its scattering effectively. Instead of following the conventional approach of designing periodic structures and applying homogenization strategies to effective material parameters or surface impedances, we address the scattering suppression problem by considering both the metasurface cover and the object behind it simultaneously. As will become evident from this study, this approach is particularly important for targets with wavelength-comparable features, where diffraction and interference play major roles in shaping the scattering cross-section. Our approach formulates scattering suppression as an inverse problem[18–23], starting from a defined objective of broadband performance and systematically identifying the metasurface configurations required to achieve it.

As a prospective use case, consider the forthcoming challenge of airspace monitoring, inspired by the vision of massive drone-based delivery systems (Fig. 1A). In urban environments, clutter issues dramatically complicate radar detection. Specifically, multipath propagation caused by reflections from obstacles such as buildings and terrain becomes a major challenge. An additional problem arises from reflections off obstacles that are inevitably present near surveillance radars and fall into their sidelobes. These obstacles, with large scattering cross-sections and positioned closer than small airborne targets, can dramatically complicate detection, and filtering them is not always straightforward, even when applying angle of arrival information. A potential solution is to cover major obstacles with scattering suppression structures. Given the complexity and diversity of different scattering scenarios, one of our objectives hereafter will be to demonstrate the advantages of the tailor-made approach. It is worth noting that this scenario is just one of many potential use cases where tailor-made metasurfaces can be applied.

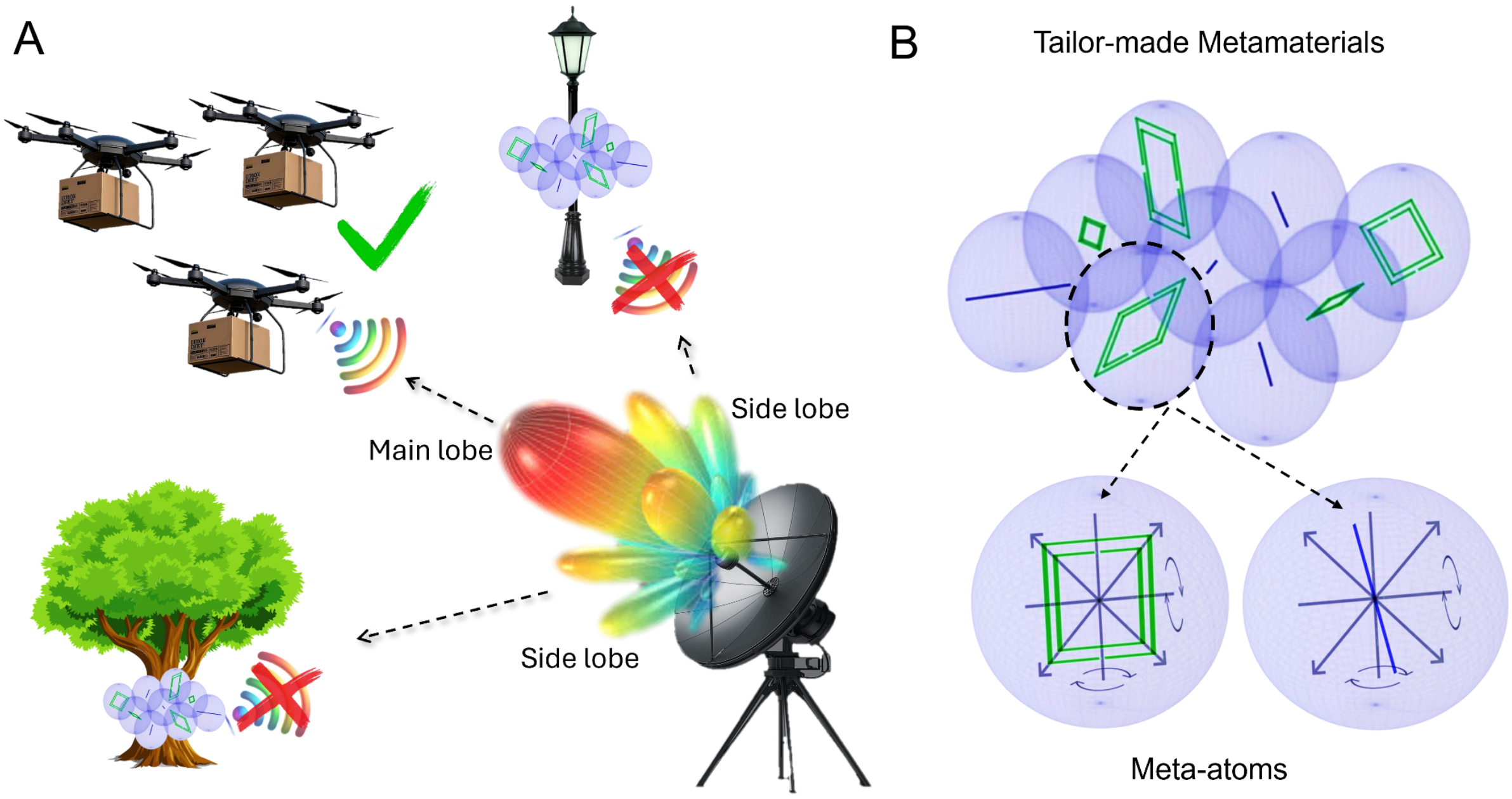

**Figure 1.** Concept of tailor-made metasurfaces for backscattering suppression. (A) A prospective scenario of airspace monitoring challenges. Reflection from obstacles positioned near a surveillance radar and falling into its sidelobes is suppressed to prevent interference and improve the accuracy of target detection. (B) The layout of a tailor-made metasurface, consisting of an array of meta-atoms, i.e., electric and magnetic resonators. In the optimization process, each element is allowed to change its size and rotate within a virtual sphere encompassing it.

## Results

This section presents the main conceptual steps and the results, while major technical details on specific realizations are provided in the Methods section.

### *The Scattering Suppression Problem*

Consider an arbitrary electromagnetic object subjected to scattering cancellation. A target may exhibit a substantial scattering cross section due to its geometry with large-scale features exceeding the wavelength, resonant properties, or a combination of both, with the latter being the most challenging case to address. While our approach is scalable and applicable across any spectral domain, we focus on the C-band (4–8 GHz) as a specific yet versatile example, given its widespread use in radar traffic control, 5G networks, and satellite communications and monitoring. As test objects, we selected a wire mesh sphere, a large random array of metal wires, and a randomly assembled polyhedron with sides composed of metal wires (Fig. 2). These structures, spanning several wavelengths, exhibit both resonant and specular reflections. While this diversity demonstrates the versatility of the method, it also captures a variety of realistic electromagnetic scenarios. For example, the wire mesh sphere is a Faraday shielding device that exhibits an omnidirectional scattering response and supports collective Mie resonances within the considered frequency range. The randomly distributed wires, while providing a broadband scattering response, are highly sensitive to the introduction of additional resonant structures in their vicinity due to near-field interactions with neighboring elements. The polyhedron serves as a generic example of a structure with a relatively high radar cross-section (RCS) exceeding 1000 $cm^2$ and was subsequently chosen for experimental demonstration. The operational bandwidth is targeted at 1-1.5 GHz, aligning with widely accepted wireless communication standards, which typically involve about 10% fractional bandwidth for radar systems. In practical terms, a 30 dB reduction in scattering corresponds to the difference between a 5 $m^2$ and 0.005 $m^2$, featuring the gap between conventional and stealth aircraft. Therefore, a 30 dB reduction, which is the target of our optimization, serves as a benchmark.

*Metasurface Optimization Search Space*

To suppress backscattering (monostatic RCS), a tailor-made metasurface is designed to be positioned in front of the structure. The architecture of the metasurface consists of an array of electric and magnetic dipoles, specifically wires and square split resonators, respectively (Fig. 1B). This configuration utilizes interactions with both fields, offering an additional layer of flexibility[24]. A specific architecture consists of a square metasurface (12×12 cm) divided into nine equal cells, where each cell represents a virtual sphere encompassing a resonator[25]. The resonator's parameters, including its overall size and spatial orientation (defined by Euler angles), are subject to optimization. The search space in this case comprises 4 continuous spatial variables and 1 discrete parameter (either electric or magnetic resonator) per cell. Consequently, the optimization effort increases substantially with the growth of array dimensions. The metasurface is positioned at a variable distance, an additional parameter, ranging from 40 mm to 200 mm below the object. Fig. 1a illustrates the layout, while Fig. 2 shows the actual tailor-made realizations.

*Evolutionary Optimization - Covariance Matrix Adaptation Evolution Strategy*

Considering the complexity of the problem, designing a tailor-made metasurface requires intensive electromagnetic optimization. This calls for selecting an appropriate optimization algorithm alongside a fast direct solver.

Optimization algorithms can be divided into two categories: gradient-based[26,27] and gradient-free methods[22,28–31]. In electromagnetic design, multi-resonant structures exhibit rapid fluctuations in their RCS due to numerous local extrema, causing instability in traditional optimization methods like grid search, steepest descent, or conjugate gradient[26]. These methods often converge to local extrema rather than the global one. Additionally, gradient-based methods are ineffective for problems with discrete components, such as here with wires and SRR mixtures, requiring combinatorial approaches. Gradient-free algorithms, such as evolutionary methods, rely on principles of natural selection, crossover, and mutation[32,33]. Starting with a random population, configurations undergo crossover and mutation to form new populations. The most adapted individuals, showing the highest objective function, are selected for the next generation. While these methods don't guarantee finding the global maximum, heuristic principles make them effective. Among these, the covariance matrix adaptation evolution strategy (CMA-ES)[34] stands out for handling high-dimensional, multi-modal problems. Unlike traditional genetic algorithms, CMA-ES efficiently explores irregular optimization landscapes, offering faster convergence and requiring minimal parameter tuning[35]. Its data-driven mutation and implicit crossover make it particularly suitable for complex electromagnetic problems.

As a forward solver, we employ the Method of Moments (MoM), specifically implemented to handle configurations composed of metal wires[36]. In this case, calculating the cost function takes about 10 to 30 seconds on an Intel® Xeon® CPU E3-1230 v6 at 3.50 GHz, enabling extensive optimization. In comparison, a similar routine would take approximately 100 times longer using commercial software such as CST Microwave Studio. Optimizing the metasurfaces described below requires 200 to 400 iterations, highlighting the need for a fast solver implementation, which reduces computation time from approximately two weeks to just three hours.

*Broadband Scattering Suppression, Examples*

The algorithm's outcomes are object-specific 3×3 arrays, with the layouts provided at Fig. 2, which summarizes the results, where three different structures are organized in rows. Panels A, D, and G show the target geometries with the designed metasurfaces beneath them. The structures are illuminated from below (the wave vector is parallel to the z-axis), with linear polarization along the x-axis. Panels B, E, and H present the backscattering spectra in $cm^2$, with the frequency ranges targeted for scattering suppression indicated by a semi-transparent green area. These ranges vary slightly to target the frequency regions where each object exhibits the strongest scattering. In these spectra, the blue dashed line corresponds to the uncovered object, while the red solid curve represents the estimated response obtained from the optimization algorithm. Since the structures exhibit multiresonant responses and may introduce instabilities, the performance was verified using the Finite

Element Method implemented in CST Microwave Studio, which accounts for the material parameters of the wires. The discrepancies between MoM and CST highlight the resonant nature of the approach and, to some extent, reflect the challenges inherent in the superscattering problem, where strong near-field accumulation within the object increases sensitivity to material losses and fabrication imperfections[36–38]. All three custom metasurfaces achieve approximately 30 dB of scattering suppression (MoM model), with the suppression reaching 40 dB for the sphere, which inherently has quite strong scattering. Notably, the bare metasurface itself exhibits significant scattering, emphasizing its strong interaction with the target (this analysis will follow). Consequently, the suppression effect arises from destructive interference.

Assessing the effectiveness of the tailor-made metasurface approach compared to other scattering suppression strategies is of interest. Several alternative methods were evaluated and are demonstrated to underperform relative to our approach, with details provided in the Methods section. For example, the straightforward method of using a tilted reflector to prevent the return of incident waves and reduce the monostatic RCS proves ineffective in this context.

While evolutionary algorithms can optimize the physical layout of designs, they often do not offer insights into the underlying physical principles that drive performance. To reveal the operational principles, 3D scattering diagrams for both the uncovered and covered targets are presented in panels C, F, and I, illustrating the physical concept of the effect: the suppression of the backscattering lobe in favor of enhanced forward scattering. Thus, the operational principle can be partially associated with first-generation stealth technology. However, unlike stealth methods based on geometrical optics, which cannot be applied to resonant objects comparable to the wavelength, our approach is multi-resonant. Furthermore, in our case, most of the energy is scattered forward, making the approach resilient to multi-static radar interrogation. The operation is preserved over a wide frequency range, as also highlighted by the scattering diagrams.

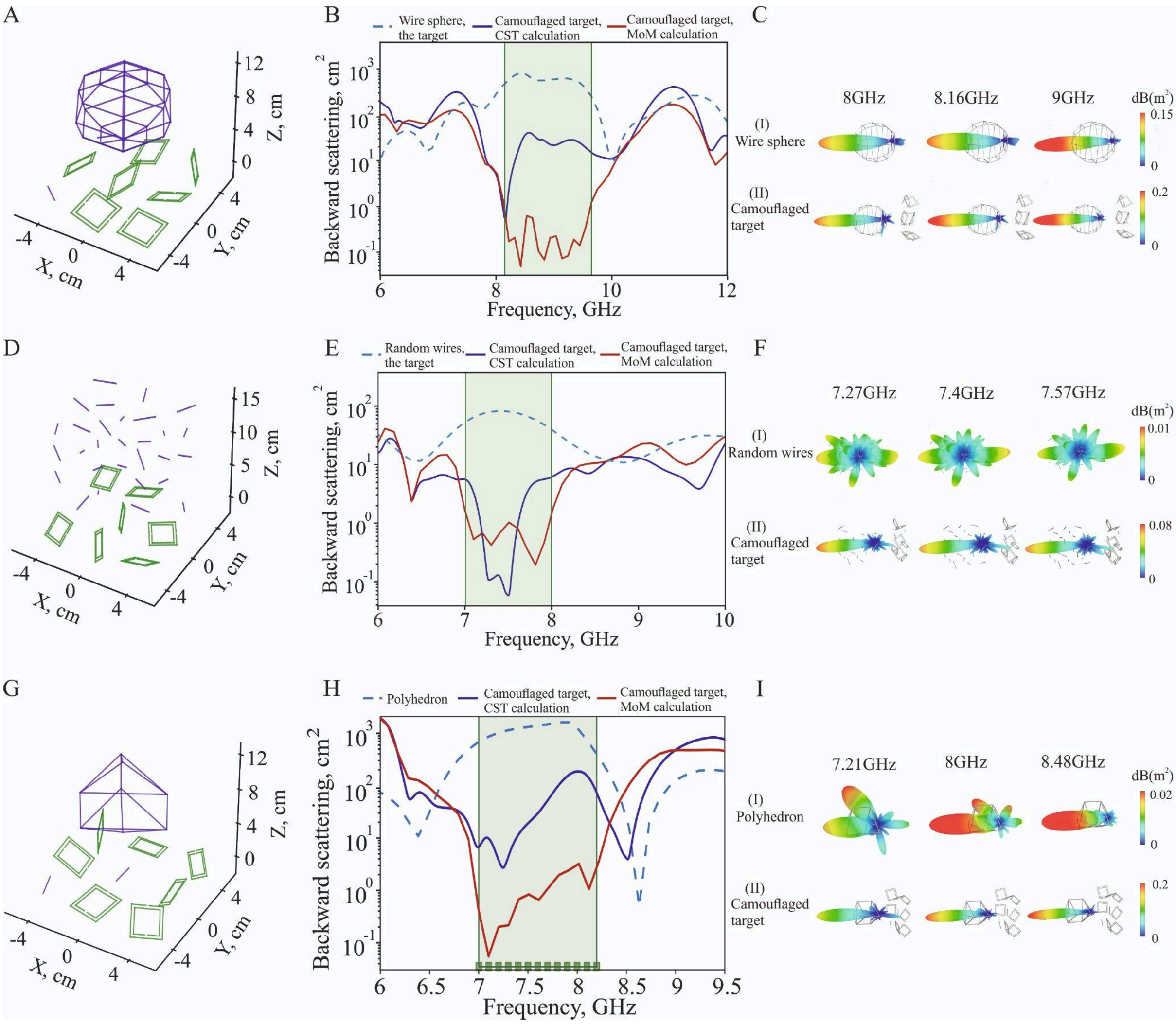

**Figure 2. Scattering analysis for three different targets:** (A–C) wire sphere, (D–F) randomly distributed dipoles, and (G–I) polyhedron. Panels A, D, and G show the layout of the target with their corresponding tailor-made metasurfaces. Panels B, E, and H are the scattering spectra, where green areas indicate the frequency ranges where the scattering suppression algorithm is applied. The light blue dashed curve represents the uncovered target, the red curve shows the camouflaged target response assessed using MoM analysis, and the blue solid curve shows the camouflaged target analyzed with CST. Panels C, F, and I present the far-field scattering patterns for covered and uncovered scenarios at representative frequencies.

*Experimental Validation on Polyhedron*

To validate the optimization approach, we conducted experimental measurements using a polyhedron. This shape was chosen because it incorporates both large-scale features, with lengths comparable to the wavelength, and resonant properties, making it a representative model of a real-world radar target. Compared to simpler geometries such as flat surfaces or spheres, the polyhedron introduces a more complex scattering pattern (Fig. 2I) that challenges conventional cloaking methods.

A photograph of the target positioned in the anechoic chamber is shown in Fig. 3A. The polyhedron is constructed from copper wires with a diameter of 1 mm. In the next stage, the tailor-made metasurface was constructed. Since scattering suppression relies on engineering multiple resonances, the experimental realization requires low-loss material platforms for resonator fabrication and accurate methods for their three-dimensional alignment. In the initial stage, the resonators forming the metasurface were fabricated using standard lithography methods. Thin, low-loss substrates were selected to minimize attenuation. Due to the complex spatial orientation and the need for precise positioning (see Fig. 3 for a sample view), a specialized holder was developed (see Methods).

Scattering measurements were conducted in an anechoic chamber, enabling accurate retrieval of absolute RCS values (Fig. 3A). The RCS was quantified using a calibration target (disc) with known parameters[39,40]. Fig. 3B summarizes the experimental results for three samples: the uncovered polyhedron, the standalone metasurface, and the concealed target. Absolute values from both the experiment and modeling agree quantitatively for the polyhedron. The concealed target also closely follows the CST model, with a slight underperformance at 8 GHz, highlighting the challenges of implementing a multiresonant approach. The dip at 7 GHz exceeds expectations and demonstrates a 35 dB suppression, indicating a slight shift in resonance interference to lower frequencies.

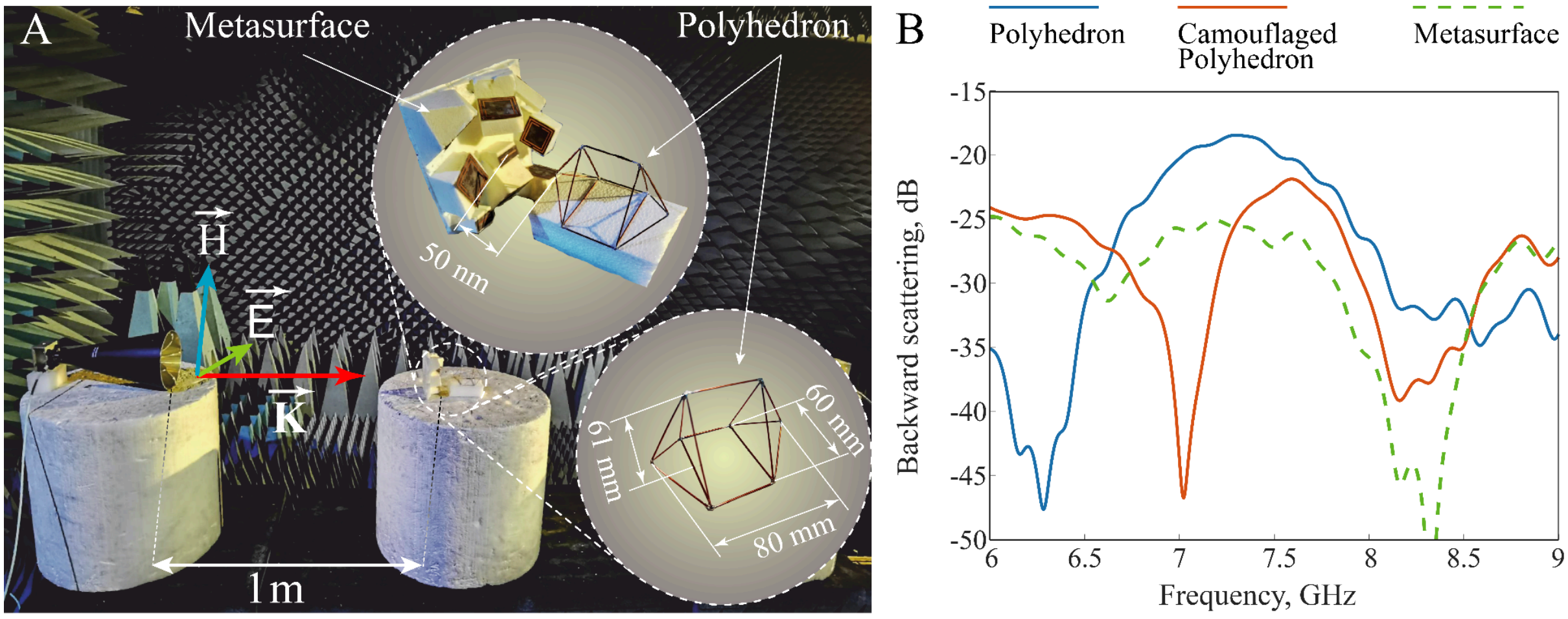

**Figure 3.** Experimental Demonstration of Tailor-Made Metasurface Camouflage. (A) Photograph of the measurement facility: an anechoic chamber with a zoom into the fabricated sample. (B) Backscattering spectra: the uncovered polyhedron (blue), the standalone metasurface (dashed green), and the concealed target (red).

## Methods

This section provides comprehensive insights and the rationale behind the undertaken steps.

*The Cost Function (Fitness Function)*

The objective (fitness) function is defined as the mean backward scattering cross-section of a metasurface-covered object, evaluated at several frequencies spanning the designated bandwidth. For clarity, the following function $F(\theta_j, R_k, \Delta_l)$ is the subject to maximization:

$$F(\theta_j, R_k, \Delta_l) = -\frac{1}{N}\sum_{m=1}^{N} \sigma(\omega_m). \quad \text{Eq. 1}$$

Here $\theta_j$ is a set of Euler angles of all resonators in the metasurface, $\{R_k\}$ a set of dimensions, characterizing wires and SSRs, $\Delta_l = \{1, 0\}^9$ a set, representing a type of resonator in each virtual sphere, and $\sigma(\omega_m)$ backward scattering at $\omega_m$ frequency. The number of frequencies ($N$), where the optimized scattering is calculated, is set manually to reach the best performance.

*CMA-ES Algorithm*

To address the high-dimensional optimization space, we used an improved variation of an evolutionary algorithm. CMA-ES (Covariance Matrix Adaptation Evolution Strategy) is a gradient-free evolutionary algorithm that utilizes the covariance matrix of the individual distribution to enhance the likelihood of finding the global minimum of a given task[34]. The strategy is effective in solving non-convex and non-smooth optimization tasks. Unlike traditional genetic algorithms (GAs), particle swarm optimization (PSO), differential evolution (DE), or simulated annealing (SA), CMA-ES is particularly well-suited for high-dimensional, multi-modal optimization problems without requiring extensive tuning[35]. Its adaptive covariance matrix allows efficient search space exploration and faster convergence. With its data-driven mutation and implicit crossover, CMA-ES is ideal for complex electromagnetic problems where the objective function landscape is highly irregular.

The concept of the designed algorithm is the following. At each step, the fitness function (Eq. 1) was evaluated for every individual in the current population, and the resulting value was assigned to that entity. The individuals were ranked from highest to lowest fitness, with the top half of the population advancing to the next iteration while the remainder was eliminated (Fig. 4).

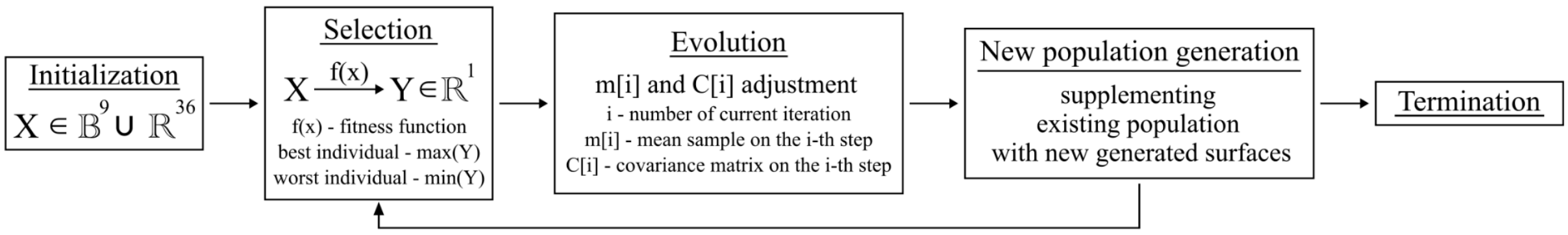

**Figure 4.** The concept of an evolutionary algorithm applied to metasurface design. The set X represents the search space of possible metasurface configurations for a given target. The space $B^9$ denotes 9-dimensional binary vectors that encode the type of resonator placed in each virtual sphere (either a wire or a square split-ring resonator). The space $R^{36}$ corresponds to 36-dimensional continuous vectors describing the geometric parameters of each resonator, including orientation (Euler angles) and size (wire lengths or square side lengths). The set $Y \subset R$ represents the continuous one-dimensional space of fitness values used to evaluate and rank the generated metasurfaces.

The following main steps of the algorithm can be highlighted. First, an initial population of metasurfaces is created, represented as 45-dimensional vectors during the initialization step (Fig. 4). The initial population, as well as each subsequent generation, is based on a normal (Gaussian) distribution from which individuals are generated:

$$x[i] = m[i] + \sigma N(0, C[i]). \qquad \text{Eq. 2}$$

Eq. 2 defines the procedure for generating new individuals in the population. Here, [i] denotes the i-th iteration, x[i] is an individual in the population, m is the population mean (center) at iteration i, σ is the step size (set to 1.3 in our algorithm), and C is the covariance matrix of the metasurface distribution at the current iteration. It is important to note that binary vectors are also generated using a normal distribution. Specifically, if the probability $P(x < a)$, where a is a randomly generated number in the range [0, 1], exceeds 0.5, the corresponding value in the binary vector is set to 1; otherwise, it is set to 0. After generating the initial population (Fig. 4), fitness values are computed for each metasurface. The population is then ranked based on these fitness values, and the top 50% of individuals (i.e., those with the highest fitness) are selected. This selected subset is used to compute updated values for m and C for the next iteration [i+1]. The update rules for m and C, as shown in Eq. 3, incorporate both the previous values and statistics computed from the selected individuals, facilitating the evolutionary optimization process. In the update of the covariance matrix C, weighting coefficients of 0.8 and 0.2 are applied to mitigate the influence of numerous local minima. The second term in the equation for C represents the outer product of the averaged selected vectors, capturing their pairwise dependencies.

$$m[i+1] = m[i] + \sigma \sum_{j=1}^{M} \frac{1}{M} x_{j,selected}^{[i]},$$ Eq. 3

$$C[i+1] = 0.8 * C[i] + 0.2 * \left(\sum_{j=1}^{M} \frac{1}{M} x_{j,selected}^{[i]}\right) \otimes \left(\sum_{j=1}^{M} \frac{1}{M} x_{j,selected}^{[i]}\right)^{T}.$$

Here, M denotes the number of top-performing metasurfaces selected for the next iteration, and $x_{j,selected}^{[i]}$ represents the j-th selected individual at iteration i. After computing the updated mean and covariance matrix, the algorithm proceeds to the next iteration, generating a new population based on these revised parameters.

*Optimization Algorithm: Proof of Concept in a Multi-Minima Landscape*

To demonstrate the capability of the algorithm to locate global minima, we constructed an artificial test function with multiple extrema. For visualization purposes, the function is defined over two variables. The positions of local minima were randomly assigned, while a global minimum was manually embedded in the landscape (corresponding to the dark-blue region, Fig. 5A). To further increase complexity, random values were added at randomly selected grid points, rendering the resulting fitness function non-differentiable. The resulting landscape is shown in Fig. 5A. Fig. 5B and Fig. 5C illustrates the initialization of the algorithm and its state after 10 iterations, clearly indicating convergence toward the region of the global minimum. In the CMA-ES algorithm, the covariance matrix defines the shape, size, and orientation of the ellipse that characterizes the sampling distribution of the population. The axis along which the ellipse is most elongated reflects the direction in which the algorithm explores most actively, potentially corresponding to regions of higher uncertainty or greater potential for improvement in the fitness function. As the optimization progresses, the ellipse tends to shrink and align with specific directions, indicating convergence toward more optimal regions of the search space. Despite the increased complexity of the objective landscape in this example, the algorithm successfully converged to the global minimum. It is important to note, however, that a similar graphical representation of the optimization process in the full 45-dimensional metasurface design space is not feasible. Therefore, convergence characteristics and performance comparisons in the metasurface setting will be presented through selected representative cases in the following sections.

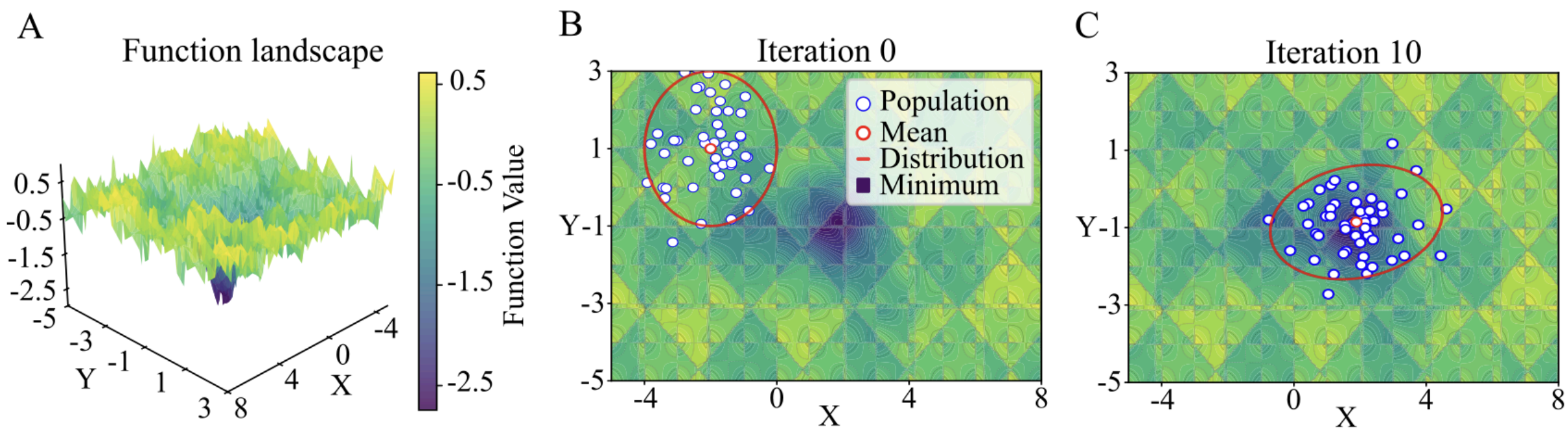

**Figure 5.** An illustration of the CMA-ES algorithm operating in a 3D search space. (A) The landscape of a non-differentiable fitness function with multiple local extrema; the dark-blue region indicates the location of the global minimum. (B, C) Snapshots of the optimization process overlaid on the function's colormap: (B) initial population, (C) after 10 iterations. White dots represent individual candidates in the population, while the red dot denotes the population mean. The red ellipse illustrates the covariance structure of the sampling distribution, indicating the principal direction and spread of the search at each stage. Ellipse is set by the equation $(x - m)^T C(x - m) = const$, where const=5.991, 95% quantile of the distribution.

*Assessment of the Algorithm Performance*

To assess the impact of the optimization on electromagnetic performance, several investigations were performed, addressing two key questions: (i) whether optimization is necessary or if similar performance could be achieved by chance from a random set of configurations, and (ii) whether the optimization search space is appropriately defined or can be reduced without degrading performance.

First, we compared the performance of the tailor-made metasurface with a randomly generated set of possible realizations. A single resonant dipole was used as the target for scattering suppression. This simplified scenario was chosen to reduce runtime and facilitate the collection of larger statistical datasets.

The test set consists of arrays that comply with the search space's constraints, where nine randomly distributed resonators have not undergone any optimization. In total, 1000 numerical experiments were conducted to collect statistical data. The realizations follow the following distribution: resonators are centered at the nodes of the symmetric array, and Euler angles are uniformly distributed. The probability of populating a cell with either a wire or a square SRR is 0.5. The resonators are positioned at the cell centers and cannot be shifted.

The optimized set was constructed by running the CMA-ES algorithm with 1000 seeds, which represents the initialization, as shown in Eq. 2. The seeds were picked randomly within the search space. The number of algorithm iterations in these experiments was fixed at 100. After collecting the data, the scattering suppression capabilities were evaluated based on the fitness function, and performance statistics for both sets were retrieved. The results, summarized in a histogram, are shown in Fig. 6A. Several assessments can be made based on different figures of merit. First, the means of the distributions can be evaluated. Both datasets exhibit distributions close to Gaussian, with the two bright vertical lines in Fig. 6A representing their respective means. The optimized metasurface exhibited a scattering suppression of 20 dB compared to the random set under this peak-to-peak assessment. However, a more relevant comparison is made by considering the best individuals in both sets. In this case, a 30–40 dB advantage is observed in favor of the optimized structure. Moreover, the best individual from the random set did not outperform the worst realization of the optimized surface. The immediate conclusions are that (i) intensive optimization is necessary, and (ii) a large set of seeds should be considered in the optimization strategy. Variations in seeds can result in differences of 20 dB or more, further demonstrating that finding the global extremum with a single seed is not guaranteed, though a properly arranged algorithm can significantly simplify the task. Thus, tens of seeds were chosen to obtain the results, as demonstrated in Fig. 2.

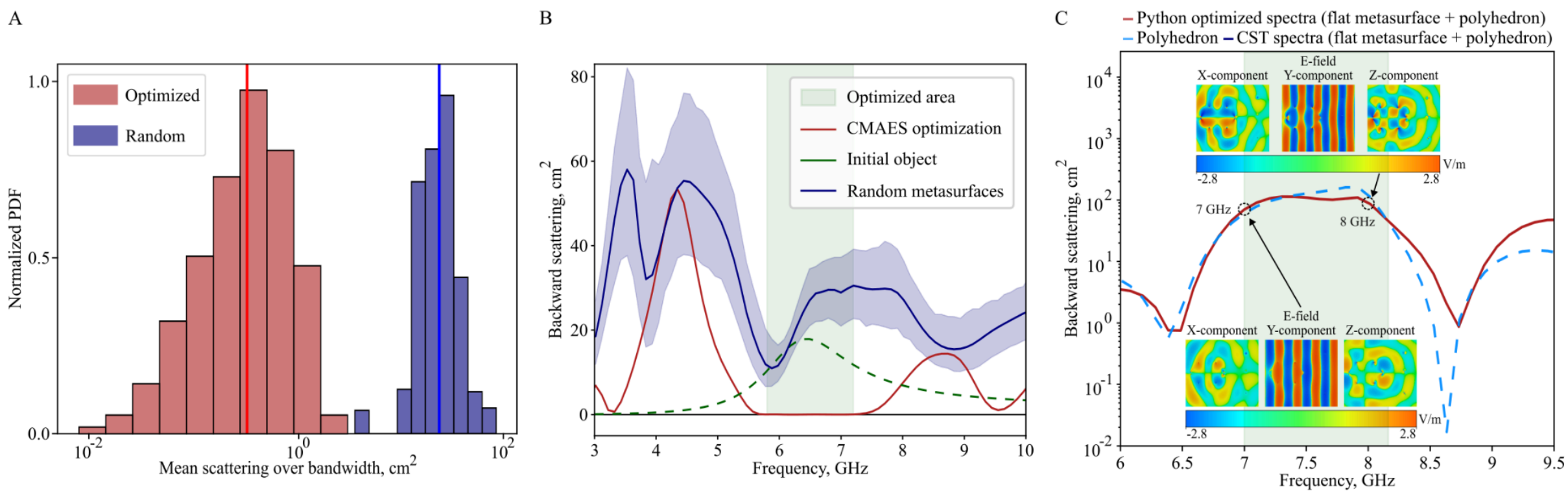

**Figure 6.** (A,B) Statistical assessment of the CMA-ES optimization algorithm for suppressing scattering from a dipole. (A) Comparison between two datasets: distribution of the fitness function for optimized metasurfaces and randomly created metasurfaces. (B) Backscattering spectra for the scenarios indicated in the legend. The test target is a single resonant dipole. (C) Attempts to obtain scattering suppression from the polyhedron using the flat optimized metasurface. Inset: The fields, as shown in the legends, demonstrate the inefficiency of this approach.

To further emphasize the comparison, the backscattering spectra were plotted in Fig. 6B. All realizations from the random set were plotted and blurred to create the blue area in the plot. The blue solid line represents the mean. The red solid curve shows the result of the best seed in the optimization. The sky-blue dashed line represents the backscattering from the initial (uncovered) dipole. The green semi-transparent bar indicates the frequency band where the optimization was performed. While CMA-ES successfully optimized the metasurface geometry for the dipole, the random structures not only failed to suppress scattering but also, in most cases, increased the backward scattering. Additionally, despite the significant broadband scattering suppression with the metasurface, new scattering peaks emerge outside the optimized region, a phenomenon that appears to be quite universal[41].

In the second part of this section, we address whether the choice of the search space is justified or if it can be simplified. Specifically, metasurface designs often comply with PCB technology, where resonators are implemented on a planar surface, which is a technologically favored solution. In our design, however, the wires and SRRs are allowed to orient out of the plane of the metasurface, raising the question of whether this feature is truly necessary or if a simpler approach could suffice.

To further investigate this, we performed the same optimization routine but restricted the resonators to move only within the plane. Figure 6C presents the results for the polyhedron target concealed with a flat structure. The scattering suppression is minimal, indicating that the two-dimensional approach is ineffective. This result arises from the fact that the target generates significant out-of-plane longitudinal fields. To suppress these fields, the metasurface resonators must be capable of generating out-of-plane currents, which can only be achieved if this additional degree of freedom is included. Figure 6C illustrates the fields that support these conclusions.

*Electromagnetic Modeling*

Scattering calculations are performed using the Method of Moments (MoM)[42], based on the Hallén integral equation, specifically adapted for thin metallic wires. This approach significantly reduces the dimensionality of the problem and computational cost compared to more computationally expensive methods such as Finite Element Methods (FEM)[43,44]. The PyNEC package[45], based on NEC-2, implements MoM and is further optimized with parallel computation over multiple CPU cores, allowing for concurrent evaluation of individuals in the population during the optimization process. While MoM-based calculations provide a highly efficient optimization tool, the final results were verified using a frequency-domain solver (FEM) implemented in CST Microwave Studio 2024. To ensure accurate results and eliminate unphysical reflections from the simulation domain boundaries, Perfectly Matched Layer (PML) boundary conditions were applied. These boundary conditions absorb outgoing waves, preventing reflections that might otherwise distort the results. The mesh resolution for all models was approximately 1–2 million cells, and the incident plane wave was polarized along the y-axis, with the wave vector directed along the z-axis (Figure 2A, D, and G). The CST modeling results aligned with those obtained using MoM, confirming the validity of MoM as the forward solver in the optimization process.

*Scattering Suppression Assessment*

In the following analysis, we focus on the polyhedron as the target for detailed investigation to reveal the electromagnetic mechanisms behind the tailored metasurface's scattering suppression operation. In this context, we evaluate another backscattering suppression strategy inspired by the first generation of stealth, where planes are designed to never be perpendicular to the incident wavevector. However, as we demonstrate here, this concept fails, at least in the context of wavelength-comparable targets.

Figure 7A presents the backscattering spectra for various scenarios. The blue solid line represents the scattering from the initial (uncovered) polyhedron. The red solid line shows the target response when concealed by the optimized metasurface, providing up to 30 dB suppression over a broad frequency range. The optimized metasurface, when detached from the target, demonstrates stronger scattering as a standalone. The comparison between the results shows that the metasurface strongly interacts with the target itself, thereby verifying the concept of the tailor-made approach. The purple-dashed line shows the response of a polyhedron covered by a tilted metal plane made of a perfect electric conductor (PEC). The plane is positioned at 35° tilt to the incident wavevector, so the Fresnel reflection law suggests that backscattering will be minimized. While this holds true for large-scale mirrors, wavelength-comparable surfaces support strong diffraction at the edges, which significantly contributes to scattering. As a result, the backscattering of the system is quite large. It is worth noting that the effect is not resonant, and the scattering cross-section remains high over the entire bandwidth. Figure 7B demonstrates the scattering diagrams for all four scenarios. The representative frequencies are shown on the plots and correspond to three characteristic frequencies, marked by crosses in panel A. This analysis further reveals the physics behind the interaction scenario. The metasurface, tailored to the target, interacts with it, reconfiguring the scattering and minimizing backscattering in favor of other directions. In this particular case, most of the energy is scattered forward.

Figure 8 (available as an animated GIF online) provides full-wave dynamic modeling of the interactions. The animation clearly shows the strong coupling between the target and the metasurface, which effectively cancels the backscattering.

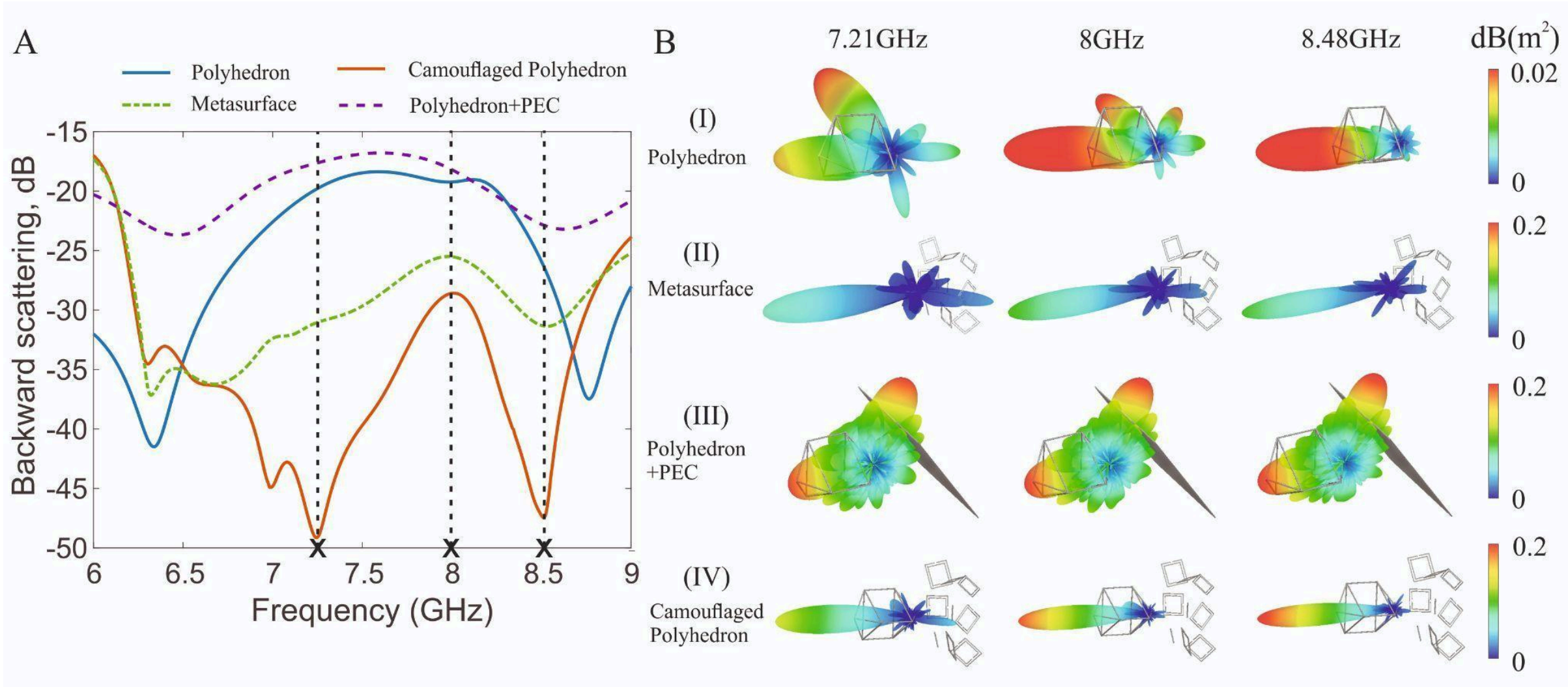

**Figure 7.** Scattering analysis of four different structures: (i) a bare polyhedron (i.e., target), (ii) the metasurface alone, (iii) a target with a PEC plate tilted at 35° in front, and (iv) a camouflaged polyhedron. (A) Scattering spectra. (B) Far-field scattering patterns for all scenarios at representative frequencies, as shown in the insets.

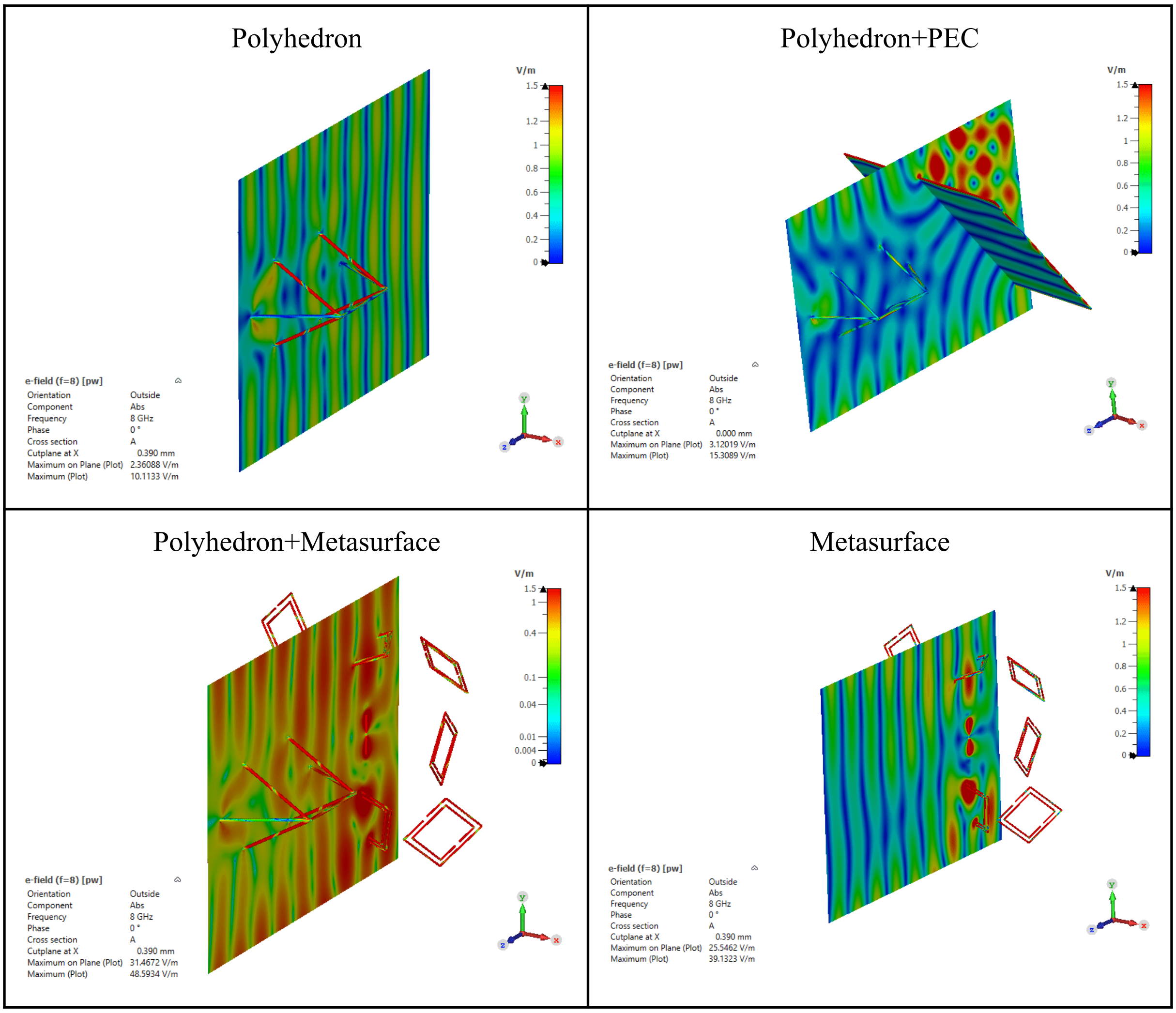

**Figure 8.** Animated images of scattering scenarios as indicated in the insets. The frequency of the incident wave is 8 GHz.

*Fabrication*

Given the highly resonant nature of the phenomenon, both material losses and fabrication imperfections in alignment must be minimized. Similar challenges arise in the design of superdirective antennas, where precise multipolar interference must be maintained[46,47]. The fabrication process consisted of two steps: (i) manufacturing individual resonators and (ii) spatially positioning the resonators in 3D space.

*The target:* The Polyhedron Scatterer is an axisymmetric pyramidal structure, specifically a convex prism with a triangular base. It is constructed from enamel-coated copper wires with a diameter of 1 mm (including an insulation thickness of 0.05 mm). The corresponding wire lengths are 40 mm and 60 mm. The wires forming the structure are soldered at the nodes to ensure electrical conductivity.

*Individual resonators:* Individual scatterers forming the metasurface were fabricated on a thin (0.15 mm) foil-clad laminate using photolithography with a photoresist and a pre-prepared mask, followed by chemical etching in a sodium tetrachlorocuprate solution. The substrate (Astra MT77, $\varepsilon_r$=3 tan(δ)=0.0017) was chosen to minimize the impact of permittivity on the resonant response. Following the design, a 3×3 array was fabricated on a thin substrate and then cut into nine pieces, each containing an individual scatterer - Figure 9A.

*RF-Transparent Holder for 3D Arrangement of Scatterers:* The main challenge in fabricating the sample is the precise alignment of resonators in space, requiring accurate tuning of three Euler angles and the height above the target for each resonator. Additionally, the material used for the holder must be transparent to GHz waves. Following these constraints, foam pillars were fabricated using the following main steps (Figure 9B and C):

1. **Model Creation:** A digital model of the pillar structure was designed based on the CST simulation, and an STL file was extracted.
2. **3D Printing:** The pillars were initially implemented by 3D printing using PLA plastic. However, 3D-printed plastic pillars cannot serve as holders, as the material itself has a background refractive index with losses [47], which would degrade performance.
3. **Silicone Mold Fabrication:** An inverted silicone mold was created using silicone rubber (Smooth-On, USA). The 3D-printed form was placed inside a box, which was then filled with silicone. Once the silicone dried, the 3D-printed structure was removed, leaving a stretchable mold.
4. **Foam Casting:** The silicone mold was lubricated with silicone grease and filled with polyurethane foam. After the foaming, expansion, and hardening processes, the holder was carefully removed from the template. Electromagnetic scattering tests confirmed its transparency, comparable to Styrofoam.
5. **Final Assembly:** The cut resonators were glued (PVA glue) onto the foam pillars, completing the metasurface assembly

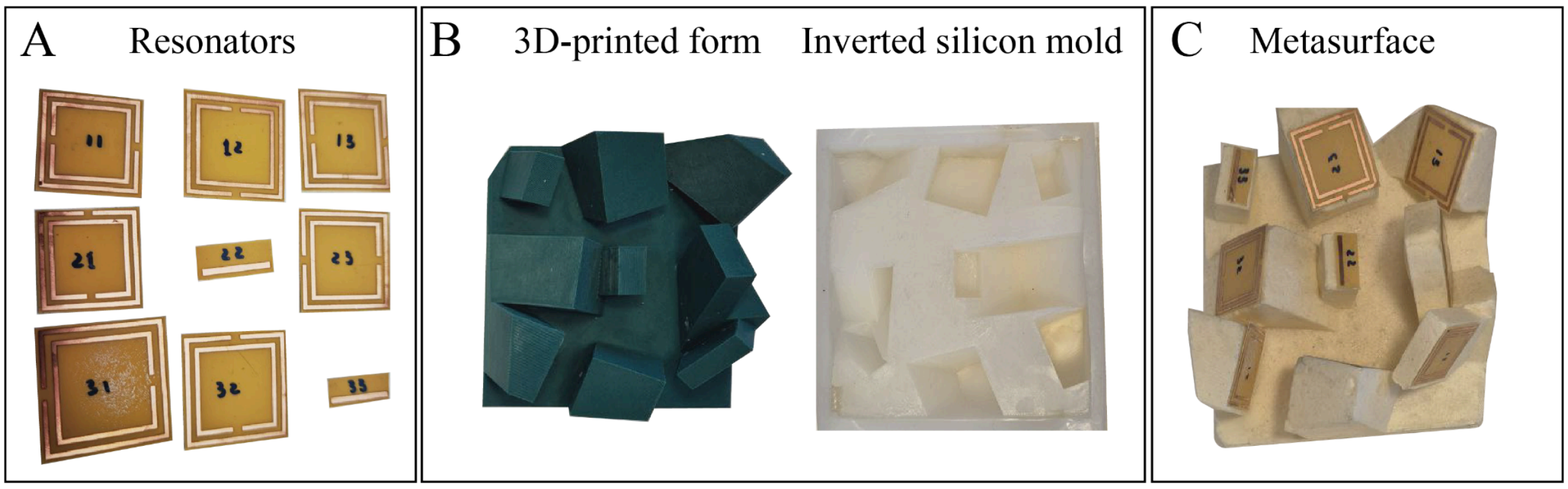

**Figure 9.** Metasurface Fabrication. (A) Resonators forming the array. (B) Photographed the main steps of the holder fabrication process. (C) Photograph of the assembled metasurface.

*Experimental Setup and RCS Measurements*

The measurements were conducted in an anechoic chamber. The experimental setup included a broadband horn antenna (NATO IDPH-2018) for both transmission (Tx) and reception (Rx), certified for the 2–18 GHz frequency range. The antenna was connected to a PNA Vector Network Analyzer (N5253B), and complex-valued S-parameters were recorded. The backscattering cross-sections of all samples were quantified using a calibration target (disc) with known parameters. This approach follows well-established and widely recognized RCS measurement technique[39].

*Conclusion*

A new concept of tailor-made metasurfaces has been proposed and demonstrated in the context of scattering suppression from arbitrary objects. Traditional metasurfaces, developed for such purposes, are typically designed with ground planes that isolate the metasurface from interacting with the object it is placed on. However, this standard approach overlooks the fine details of the target, which can significantly impact performance. The problem becomes more pronounced when the object exhibits complex shapes and small features that do not align with the application of metal substrates. Additionally, when the target is comparable in size to the wavelength, traditional large-scale metasurfaces, with unit cell periods related to the wavelength, are no longer applicable. In these cases, the periodic nature of the metasurface becomes a limiting factor. Furthermore, diffraction and scattering effects at the edges become more pronounced, leading to performance degradation. Our approach addresses these challenges by designing metasurfaces that are specifically tailored to the object, enabling superior suppression of scattering, even in the presence of complex geometries or wavelength-comparable objects.

The tailor-made metasurface architecture consisted of a 3x3 array, where the basic elements are either electric (wires) or magnetic (SRR) resonators, packed together. The strong magneto-electric interactions within the array allow it to interact with the near-field of the target, creating a strong collective response. Given the complexity of the optimization problem, which aims to achieve broadband and resonance-based backscattering suppression simultaneously, the Covariance Matrix Adaptation Evolution Strategy (CMA-ES) algorithm was developed and applied across a range of diverse test targets. Specifically, three main targets representing different prospective applications were tested: a randomly shaped polyhedron, a wire mesh sphere, and a large array of randomly distributed resonant dipoles. For each target, a tailor-made metasurface was designed to achieve scattering suppression of 20-30 dB over a fractional bandwidth of 10-20%, depending on the specific realization. It is worth noting that this level of performance aligns with technological standards where similar results are targeted across a variety of use cases involving wireless sensing and communications. The basic physical concept underlying the phenomenon is the resonant broadband redistribution of electromagnetic waves, suppressing backreflection in favor of forward scattering.

In a broader sense, the proposed tailor-made approach can be applied to a wide range of problems and use cases by adjusting the fitness function in the optimization tool accordingly. In the realm of wireless communication and sensing in overcrowded environments, tailoring scattering becomes a priority. The multipath problems become severe, affecting signal integrity, data rates, and the reliability of communication systems. By minimizing backscattering and redirecting energy toward favorable directions, the tailored metasurfaces can mitigate interference caused by multipath propagation. This capability is particularly beneficial in urban environments with dense buildings and obstacles, where traditional methods of signal management fail to maintain efficient communication. Furthermore, in radar systems, where detecting objects with minimal interference is crucial, our approach offers a viable solution to enhance detection accuracy and reduce clutter.